\documentclass[12pt]{article}   

\begin{document}
\title{Noncommutative space-time models }
\author{N.A. Gromov\footnote{e-mail: gromov@dm.komisc.ru},  V.V. Kuratov \\
Department of Mathematics, Komi Science Center,\\ Ural Division, Russian Academy of Sciences,\\ Chernova st., 3a, Syktyvkar, 167982, Russia}

\maketitle

\begin{abstract}
The FRT quantum   Euclidean  spaces $O_q^N$ are formulated in terms of Cartesian generators. The quantum analogs of N-dimensional 
Cayley-Klein spaces are obtained  by contractions and analytical continuations. Noncommutative constant curvature spaces are introduced
as a spheres in the quantum Cayley-Klein spaces.  For $N=5$ part of them are interpreted as the noncommutative analogs of (1+3) space-time models. As a result the quantum (anti) de Sitter, Newton, Galilei kinematics with the fundamental length and the fundamental time are suggested.
\end{abstract}

\section{Introduction}

Space-time is a fundamental conception which underlines the most significant  physical theories. Therefore  analysis of a possible space-time models (or kinematics) has the fundamental  meaning  for physics.
 Possible commutative kinematics
were described in \cite{B-68} on the level of Lie algebras.
From the point of view of geometry these  kinematics  are realized 
as  constant curvature spaces, which can be obtained from the spherical space by contractions and analytical continuations  known as Cayley-Klein (CK) scheme \cite{G-90}.

New possibility for construction of the noncommutative space-time models
is provided by quantum groups and quantum vector spaces \cite{FRT}.
The quantum Poincar\'e group related to the $\kappa$-Poincar\'e algebra
as well as the $\kappa$-Minkowski kinematics were suggested 
\cite{Z-94}--\cite{LLM}.
A general formalism that allows the construction of  field theory
in $\kappa$-Minkowski space-time was developed \cite{W-04}.
An approach  connected with quantum deformation of Lie algebras is mainly used in these papers.

The purpose of our  paper is to obtain the noncommutative (quantum) analogs of the possible kinematics
 starting with the quantum Euclidean space.  
CK scheme of contractions and analytical continuations was developed in Cartesian basis whereas  the standard quantum group theory
\cite{FRT} was formulated in a different skew-symmetric one.
Therefore first of all this theory is reformulated in the Cartesian basis,
then  the noncommutative analogs of constant curvature spaces (CCS) including fiber (or flag) spaces are investigated 
and some of them are interpreted as noncommutative kinematics. 

\section{Commutative kinematics}       
    
  Classical four-dimensional space-time models  can be obtained
 \cite{G-90} by the physical interpretation
of the orthogonal  coordinates of the most symmetric spaces,
namely constant curvature spaces. All $3^N$ $N$-dimensional CCS
are realized on the spheres
\begin{equation}
S_N(j)=\{ \xi_1^2+j_1^2\xi_2^2+ \ldots +(1,N+1)^2\xi_{N+1}^2=1 \},
\label{1}
\end{equation}
where 
\begin{equation}
(i,k)=\prod^{\max(i,k)-1}_{l=\min(i,k)}j_l, \quad
(k,k)\equiv 1,
\label{2}
\end{equation}
and each of parameters  $j_k$ takes the values $1,\iota_k,i, \  k=1, \ldots, N.$  
Here $ {\iota}_k $ are nilpotent generators
 $  {\iota}_k^2=0,  $ with commutative law of multiplication
$ {\iota}_k{\iota}_m={\iota}_m{\iota}_k \not =0, \  k \neq m. $

The intrinsic Beltrami coordinates
$x_k=\xi_{k+1}\xi_1^{-1},\; k=1,2,\ldots,N $
present the  coordinate system in CCS, which coordinate lines
$x_k=const$ are geodesic. CCS has positive curvature for $j_1=1,$ 
negative for $j_1=i$ and it is flat for  $j_1=\iota_1.$
For a flat space the Beltrami coordinates  coincide with the Cartesian ones.
Nilpotent values $j_k=\iota_k,\; k>1$ correspond to a fiber (flag) spaces 
and imaginary values $j_k=i$ correspond to pseudo-Riemannian spaces.

Classical $(1+3)$ kinematics \cite{B-68} are obtained from CCS for
$N=4, \; j_1=1,\iota_1,i, \; j_2=\iota_2,i,\; j_3=j_4=1 $
if one interprets $x_1$ as the time axis $t=\xi_2\xi_1^{-1}$ and the rest
as the space axes $r_k=\xi_{k+2}\xi_1^{-1}, k=1,2,3.$
The standard  de Sitter kinematics $S_4^{(-)}$  with constant negative curvature is realized for $ j_1=j_2=i,$ anti de Sitter kinematics $S_4^{(+)}$
with positive  curvature --- for $ j_1=1, \;j_2=i.$ Relativistic flat Minkowski kinematics $M_4$  appears for $ j_1=\iota_1, \; j_2=i.$ Nonrelativistic Newton $N_4^{(\pm)}$ and Galilei $G_4$ kinematics  correspond to $ j_2=\iota_2,$  $j_1=1,i$ and $ j_1=\iota_1,$ 
respectively.

\section{ Quantum Cayley-Klein spaces } 
Let us remind the definition of the quantum vector space \cite{FRT}.
An algebra  $ O_q^N({\bf C}) $ with generators $ x_1,\ldots,x_N $ and
commutation relations
\begin{equation}
\hat R_q(x \otimes x)=qx \otimes x-{{q-q^{-1}} \over {1+q^{N-2}}}
x^tCx  W_q,
\label{6}
\end{equation}
where
$ \hat R_q=PR_q, \; Pu \otimes v=v \otimes u,\;\forall u,v \in {\bf C}^n,\;
W_q=\sum^N_{i=1}q^{\rho_{i'}}e_i \otimes e_{i'},$ 
\begin{equation}  
x^tCx=\sum^N_{i,j=1}x_iC_{ij}x_j=
\epsilon x_{n+1}^2+
\sum_{k=1}^n\left(
q^{-\rho_k}x_kx_{k'}+q^{\rho_k}x_{k'}x_k\right), 
\label{7}
\end{equation}
$\epsilon =1 $ for $N=2n+1,$ $\epsilon =0 $ for $N=2n$
and vector $ (e_i)_k = \delta_{ik}, \enskip i,k=1,\ldots,N $
is called the algebra of functions on $N$-dimensional quantum
Euclidean space (or simply the quantum Euclidean space) $ O_q^N({\bf C}). $
         
The matrix $C$ has non-zero elements only on the secondary diagonal. They are equal to unit in the commutative limit $q=1$. Therefore   the quantum vector space $ O_q^N({\bf C}) $ is described
 in a skew-symmetric basis, where for $q=1$ the invariant form $inv=x^tC_0x$  is given by the matrix $C_0$ with the only unit non-zero elements  on the secondary diagonal. 

In the case of kinematics, the most natural basis is the Cartesian basis, where the invariant form $inv=y^ty$ is given by the unit matrix $I.$
The transformation from the skew-symmetric  basis $x$
to the Cartesian basis $y$ is described by $y=D^{-1}x,$ where  matrix $D$  is a 
solution of the  equation
$ D^tC_0D=I.$
This equation has many solutions. Take one of these, namely
\begin{equation}
D=\frac{1}{\sqrt{2}}
\left ( \begin{array}{ccc}
      I & 0 &  -i{\tilde C_0} \\
      0 & \sqrt{2} &  0 \\
      {\tilde C_0} & 0  &  iI
      \end{array} \right ),    \       N=2n+1,
\label{13}
\end{equation}
where $ {\tilde C_0} $ is the  $ n \times n $ matrix  with real units
on the secondary diagonal and all other elements equal to zero. For $N=2n$ the matrix $D$ is given by (\ref{13})
without the middle column and row. 
The matrix (\ref{13}) provides one of the possible combinations of the quantum group structure and the CK scheme of group contractions.
All other similar combinations  are given
by  the matrices $ D_{\sigma}=DV_{\sigma}, $  obtained from
(\ref{13}) by the right multiplication on the matrix $ V_{\sigma} \in M_{N} $ with elements
$ (V_{\sigma})_{ik}= \delta_{\sigma_{i},k}, $
where $ \sigma \in S(N) $ is a permutation of the $ N$-th order.

We derive the quantum Cayley-Klein spaces with the same  transformation
of the Cartesian generators
$ y =\psi \xi, \; \psi ={\rm diag}(1,(1,2),\ldots,(1,N)) \in M_N, $
as in commutative case \cite{G-90}.
The transformation  $z=Jv$ of the deformation parameter $q=e^z$ should be added in quantum case. The commutation relations of the Cartesian generators of the quantum  $N$-dimensional Cayley-Klein space are given by the equations  
$$
{{\hat R}}_{\sigma}(j) \xi \otimes \xi = e^{Jv} \xi \otimes \xi -
{{2sh Jv} \over {1+e^{Jv(N-2)}}}\xi^tC_{\sigma}(j) \xi  {W}_{\sigma}(j),
$$
where
$$
{\hat R}_{\sigma}(j)={\Psi}^{-1}(D_{\sigma} \otimes D_{\sigma})^{-1}{\hat R_q}(D_{\sigma} \otimes D_{\sigma})\Psi, \quad
W_{\sigma}(j)={\Psi}^{-1}(D_{\sigma} \otimes D_{\sigma})^{-1}{W}_q,
$$
\begin{equation}
C_{\sigma}(j)=
{\psi} D^t_{\sigma} C D_{\sigma} \psi =
{\psi} V^{t}_{\sigma} D^t C D V_{\sigma} \psi,\;\;
\Psi = \psi \otimes \psi.
\label{14}
\end{equation}
The explicit form of commutation relations see \cite{GK-04}.
The multiplier $J$  is chosen as
$ J=\displaystyle{\bigcup^n_{k=1}(\sigma_k,\sigma_{k'})}.$
This is the minimal multiplier, which guarantees 
the existence of the Hopf algebra structure for the associated quantum group $SO_v(N;j;\sigma).$ 
The ``union''
$(\sigma_k,\sigma_{p}) \bigcup  (\sigma_m,\sigma_{r})$
is understood as the first power multiplication of all parameters
$j_k,$ which occur  at least in one multiplier
$(\sigma_k,\sigma_{p})  $ or $(\sigma_m,\sigma_{r}),  $
for example, $(j_1j_2) \bigcup (j_2j_3)=j_1j_2j_3.$

Quantum orthogonal Cayley-Klein sphere $S_v^{(N-1)}(j;\sigma) $ is obtained
as the quotient of $O_v^N(j;\sigma) $ by $ inv(j)=\xi^tC_{\sigma}(j) \xi  =1. $ 
The quantum analogs of the intrinsic Beltrami coordinates on this sphere
are given by  the sets of independent right or left generators
\begin{equation}
r_{\sigma_{i}-1}=\xi_{\sigma_i} \xi_1^{-1}, \quad
\hat{r}_{\sigma_{i}-1}=\xi_1^{-1} \xi_{\sigma_i} ,
\quad i=1,\ldots ,N, \quad i\neq k, \quad \sigma_k =1.
\label{17}
\end{equation}

In the case of quantum Euclidean  spaces $ O_q^N({\bf C}) $ the use
of different $D_{\sigma}$ for $\sigma\in S(N)$ makes no sense, because
all similarly obtained quantum  spaces are isomorphic. However  
the situation  is radically different for the quantum Cayley-Klein spaces. In this case the Cartesian generators 
$ \xi_k$ are multiplied by $(1,k)$ and for nilpotent values of all or
some parameters $j_k$ this  isomorphism of quantum vector spaces is  destroyed.
 The necessity of using different $ D_{\sigma} $  arises
as well if there is some physical interpretation of generators.
In this case  physically different generators may be confused by
permutations $ \sigma, $ for example, time and space generators of kinematics.
Mathematically isomorphic kinematics may be physically nonequivalent.

\section{ Quantum kinematics} 

 For $N=5$ the thorough analysis of the multiplier
$J=(\sigma_1,\sigma_5)\cup (\sigma_2,\sigma_4),$
which  appears in the transformation of the deformation parameter
$z=Jv,$ and commutation relations                                                      (\ref{14}) 
of the quantum  space generators for different permutations allowed  to find
two permutations giving a different $J$ and a physically nonequivalent
kinematics, namely 
 $ \sigma_0 =(1,2,3,4,5)$ and $\sigma' =(1,4,3,5,2). $

In order  to clarify the relation with the standard
Inonu--Wigner contraction procedure \cite{IW-53},
the mathematical parameter $j_1$
is replaced by  the physical one  $\tilde{j}_1T^{-1},$ 
and the  parameter  $j_2$ is replaced by   $ic^{-1},$ where $\tilde{j}_1=1,i.$ 
The limit $T \rightarrow \infty $ corresponds to the contraction  $j_1=\iota_1,$ and the limit $c \rightarrow \infty $ corresponds to
$j_2=\iota_2.$ The parameter $T$ is interpreted as
the curvature radius and has the  physical dimension of time
$[T]=[\mbox{time}],$ the parameter  $c$ is the light velocity  
$[c]=[\mbox{length}][\mbox{time}]^{-1}. $

As far as the generator $\xi_1$ does not commute with others,
it is  convenient to introduce right and left time 
     $t=\xi_2\xi_1^{-1},\; \hat{t}=\xi_1^{-1}\xi_2$ and space
     $r_k=\xi_{k+2}\xi_1^{-1},\; \hat{r}_k=\xi_1^{-1}\xi_{k+2},\; k=1,2,3$
generators. The reason for this  definition is the simplification of expressions for commutation relations of          
quantum kinematics. 
The commutation relations of the independent generators are obtained
(see \cite{GK-04} for details)  in the form
$$
S_v^{4(\pm)}(\sigma_0)=\{t,{\bf r}|\;\;
\hat{t}r_1=\hat{r}_1t \cos {\frac{\tilde{j}_1v}{cT}}
+i \hat{r}_1r_2 {\frac{1}{c}} \sin {\frac{\tilde{j}_1v}{cT}}, \;
$$
$$
\hat{t}r_2 - \hat{r}_2t= -2i\hat{r}_1r_1{\frac{1}{c}}  
\sin {\frac{\tilde{j}_1v}{2cT}},\;\;
\hat{t}r_3=\hat{r}_3t \cos {\frac{\tilde{j}_1v}{cT}}
-i t {\frac{cT}{\tilde{j}_1}} \sin {\frac{\tilde{j}_1v}{cT}}, \;
$$
\begin{equation}
\hat{r}_1r_2=\hat{r}_2r_1 \cos {\frac{\tilde{j}_1v}{cT}}
-i \hat{t}r_1 c \sin {\frac{\tilde{j}_1v}{cT}}, \;\;
\hat{r}_pr_3=\hat{r}_3r_p \cos {\frac{\tilde{j}_1v}{cT}}
-i r_p {\frac{cT}{\tilde{j}_1}} \sin {\frac{\tilde{j}_1v}{cT}} \},
\label{18}
\end{equation}
$$
S_v^{4(\pm)}(\sigma')=\{t,{\bf r}|\;\;
\hat{r}_kt=\hat{t}r_k \cosh {\frac{\tilde{j}_1v}{T}}
-i r_k {\frac{T}{\tilde{j}_1}} \sinh {\frac{\tilde{j}_1v}{T}},
$$
$$
\hat{r}_2r_1=\hat{r}_1r_2 \cosh {\frac{\tilde{j}_1v}{T}}
-i\hat{r}_1 r_3  \sinh {\frac{\tilde{j}_1v}{T}}, \;\;
\hat{r}_1r_3=\hat{r}_3r_1 \cosh {\frac{\tilde{j}_1v}{T}}
-i\hat{r}_2 r_1  \sinh {\frac{\tilde{j}_1v}{T}}, \;\;
$$
\begin{equation}
\hat{r}_2r_3-\hat{r}_3r_2=2i\hat{r}_1r_1\sinh {\frac{\tilde{j}_1v}{2T}}  \}.
\label{20}
\end{equation}

In the case of the identical permutation $ \sigma_0, $ 
deformation parameter $v$ for the system units, where $\hbar=1,$
 has the physical dimension of length $[v]=[cT]=[\mbox{length}] $ 
and may be interpreted as the fundamental length.
For the  permutation $ \sigma', $ the quantum (anti) de Sitter kinematics  (\ref{20}) are characterized
by the fundamental time $[v]=[\mbox{time}]. $  
Recall that the same physical dimensions of the deformation parameter
have been obtained   for the quantum algebras $so_v(3;j;\sigma)$ and corresponding $(1+1)$ kinematics for a different permutations \cite{G-95}.

In the zero curvature limit $ T \rightarrow \infty $ two quantum
Minkowski kinematics   are obtained 
$$
M_v^4(\sigma_0)=\{t,{\bf r}|\;\;  [t,r_p]=0, \;[r_3,t]=ivt, \; 
[r_2,r_1]=0,\; [r_3,r_p]=ivr_p, \;p=1,2,    \},
$$
\begin{equation}
M_v^4(\sigma')=\{t,{\bf r}|\;\;  [t,r_k]=ivr_k, \;  \;
[r_i,r_k]=0, \;i,k=1,2,3    \}.
\label{24}
\end{equation}
The first one is isomorphic to the tachyonic $\kappa$-Minkowski kinematics,
 the second one to the standard  $\kappa$-deformation 
 \cite{Z-94}--\cite{LLM}.
For both $\kappa$-Minkowski kinematics  in the system units $\hbar=c=1$ the deformation parameter $\Lambda=\kappa^{-1}$ has
the physical dimension of length and is interpreted as the fundamental
length. But in the system units $\hbar=1$ the deformation parameter
has  different dimensions, namely $v$ is the fundamental length for
$M_v^4(\sigma_0)$  and $v$ is the fundamental time for
$M_v^4(\sigma'). $

As far as the commutation relations (\ref{24}) do not depend on $c,$
they do not change in the limit $ c \rightarrow \infty,$ therefore
the generators of the quantum Galilei kinematics $G_v^4(\sigma_0)$  and $G_v^4(\sigma')$  are subject of the same
commutation relations. 

In the nonrelativistic limit $ c \rightarrow \infty $
there are two noncommutative analogs of the  Newton
kinematics $(p=1,2) $
$$
N_v^{4(\pm)}(\sigma_0)=\{t,{\bf r} | \; [t,r_p]=0, \;\;
[r_3,t]=ivt(1+\tilde{j}_1^2{\frac{t^2}{T^2}}), \;\;
$$        
$$
[r_1,r_2]=0,\;\;
[r_3,r_p]=ivr_p(1+\tilde{j}_1^2{\frac{t^2}{T^2}})  \}, 
$$
$$
N_v^{4(\pm)}(\sigma')=\{t,{\bf r} | \; 
[t,r_k]=i(r_k+{\frac{\tilde{j}_1^2}{T^2}}tr_kt){\frac{T}{\tilde{j}_1}} 
\tanh{\frac{\tilde{j}_1v}{T}}, 
$$
$$
r_2r_1=r_1r_2 \cosh {\frac{\tilde{j}_1v}{T}}
-ir_1 r_3  \sinh {\frac{\tilde{j}_1v}{T}}, \;\;
$$
\begin{equation}
r_1r_3=r_3r_1 \cosh {\frac{\tilde{j}_1v}{T}}
-ir_2 r_1  \sinh {\frac{\tilde{j}_1v}{T}}, \;\;
[r_2,r_3]=2ir_1^2\sinh {\frac{\tilde{j}_1v}{2T}}  \},
\label{25}
\end{equation}
where in the last case the deformation parameter is not transformed
under contraction. The multiplier $T^{-1}$  appears as the result of
the physical interpretation of the quantum space generators.
For nonzero curvature kinematics commutation relations of generators  depend on $c$ and are different for relativistic and nonrelativistic cases, unlike Minkowski and Galilei kinematics.

\section{Conclusion}

We have reformulated the quantum Euclidean space $O_q^N$ in Cartesian coordinates and 
then used  the standard trick with real, complex, and 
dual  numbers in order to define the quantum Cayley-Klein spaces  $O_q^N(j;\sigma)$ uniformly. 
Noncommutative constant curvature spaces are generated by the Beltrami coordinates on spheres $S_v^{N-1}(j;\sigma).$ 
The different combinations of quantum structure
and CK scheme  are described with the help of permutations $\sigma. $ 
As a result for $N=5$, the   quantum deformations of (anti) de Sitter,
Minkowski, Newton and  Galilei  kinematics are obtained.
We have found two types of the noncommutative  space-time models with fundamental length and fundamental time.

The quantum Galilei kinematics   have  the same commutation relations  as the quantum Minkowski
kinematics.  In other words,
the quantum deformations of the flat kinematics are identical, whereas for nonzero curvature kinematics commutation relations of generators   are different for relativistic and nonrelativistic cases.

Noncommutative kinematics are obtained by the interpretation of some mathematical constructions associated with quantum groups and quantum spaces. Noncommutativity of space and time generators appear at the distance comparable with the fundamental length or at the time interval comparable with the fundamental time.
The  deformation parameter is free parameter of these models.
Which type of the model is more appropriate and what is the value of deformation parameter, i.e. the values of fundamental length and fundamental time, are questions   of experimental study. 
 
This work was supported by Russian Foundation for Basic
Research under Project 04-01-96001.

\end {document}